\documentclass{PoS}
\usepackage{cite}

\title{{\footnotesize DESY 09-011 
\\
\vspace*{-4mm}
                      SFB-CPP-09/17}\\
From Moments to Functions in Quantum Chromodynamics}

\ShortTitle{From Moments to Functions}

\author{\speaker{Johannes Bl\"umlein}\thanks{This project was supported in part by
German DFG SFB TR-9, Studienstiftung des Deutschen Volkes,
projects P19462-N18, P20162-N18, and P20347-N18 of the Austrian FWF, and
EC MRTN HEPTOOLS, contract MRTN-CT-2006-035505.}\\
        Deutsches Elektronen-Synchrotron, DESY, \\ Platanenallee 6, D-15738 Zeuthen,
        Germany\\
        E-mail: \email{Johannes.Bluemlein@desy.de}}

\author{Manuel Kauers\\
        Institute for Symbolic Computation (RISC),\\
        Johannes Kepler University, Altenberger Stra\ss{}e 69, A-4040 Linz, Austria\\
        E-mail: \email{Manuel.Kauers@risc.uni-linz.ac.at}}

\author{Sebastian Klein\\
        Deutsches Elektronen-Synchrotron, DESY, \\ Platanenallee 6, D-15738 Zeuthen,
        Germany\\
        E-mail: \email{Sebastian.Klein@desy.de}}

\author{Carsten Schneider\\
        Institute for Symbolic Computation (RISC),\\
        Johannes Kepler University, Altenberger Stra\ss{}e 69, A-4040 Linz, Austria\\
        E-mail: \email{cschneid@risc.uni-linz.ac.at}}

\abstract{Single-scale quantities, like the QCD anomalous dimensions and Wilson
          coefficients, obey difference equations. Therefore their analytic form
          can be determined from a finite number of moments. We demonstrate this
          in an explicit calculation by establishing and solving large scale
          recursions by means of computer algebra for the anomalous dimensions and
          Wilson coefficients in unpolarized deeply inelastic scattering from their
          Mellin moments to 3-loop order.}

\FullConference{XII Advanced Computing and Analysis Techniques in Physics Research\\
		 November 3--7 2008\\
		 Erice, Italy}

\begin{document}

\section{Introduction}
\noindent
Higher order calculations in Quantum Field Theories easily
become tedious due to the larger number of terms emerging and the
sophisticated form of the contributing Feynman parameter integrals.
This applies already to \emph{zero scale} and \emph{single scale}
quantities. Even more this is the case for problems containing more than
one scale. While in the latter case the mathematical structure
of the solution of the Feynman integrals is widely unknown, it
is explored to a certain extent for zero-  and single scale
quantities. Zero scale quantities emerge as the expansion coefficients
of the running couplings and masses, as fixed moments of splitting
functions, etc. They can be expressed by rational numbers and
certain special numbers as \emph{multiple zeta-values (MZVs)}~\cite{LIS,BBV}
and
related quantities.

{Single scale} quantities depend on a scale $z$ which may be
given as a ratio of Lorentz invariants $s'/s$ in the respective
physical problem. One may perform a \emph{Mellin transform}
over $z$
\begin{eqnarray}
\int_0^1 dz~z^{N-1} f(z) = M[f](N)~.
\nonumber
\end{eqnarray}
All subsequent calculations are then carried out in Mellin space
and one assumes {$N \in {\bf N},~N > 0$}. By this transformation the
problem at hand becomes {discrete}. One may seek a description in
terms of {difference equations}. {Zero scale} problems are obtained from
{single scale} problems treating {{$N$}} as a fixed integer or
considering the limit {$N \rightarrow \infty$}.

A main question concerning zero scale quantities is:
Do the corresponding Feynman integrals {always} lead to MZVs? In the
lower orders this is the case. However, starting at some
order, even for single-mass problems, other special numbers will occur
\cite{ANDRE}. This makes it difficult to use methods like {\tt PSLQ} \cite{PSLQ}
to determine the analytic structure of the corresponding terms even
if one may calculate them numerically at high enough
precision since one has to known the respective basis completely.~\footnote{In a recent
analysis \cite{BBV} the relations between all MZVs in the non-alternating and alternating
case were determined
up  to weight $w = 12$ and up to $w = 24$ in the non-alternating case
using shuffle-, stuffle- and generalized doubling relations. These relations
lead to basis lengths according to the conjectures by Broadhurst \cite{BROAD96}
and Broadhurst-Kreimer-Zagier \cite{BKZ}, which meets the upper bound set by
Terasoma, Goncharov and Deligne~\cite{TERAS}.
In the non-alternating case we verified that the latter bound is valid
at least to $w = 26$ and
stopped the calculation afterwards due to the large complexity involved. All this does
not exclude the existence of exotic relations reducing the basis further.}

Zero scale problems are much easier to calculate than single scale problems. In
some analogy to the determination of the analytic structure in zero scale problems
through integer relations over a known basis ({\tt PSLQ}) one may think of an
automated reconstruction of the all--$N$ relation out of a \emph{finite number} of
Mellin moments given in analytic form. This is possible for recurrent quantities. At
least up to {3-loop order}, presumably even to higher orders, single scale quantities
belong to this class. Here we report on a general algorithm for this purpose, which we
applied to the problem being currently the most sophisticated one: the determination
of the anomalous dimensions and Wilson coefficients to 3--loop order for
unpolarized
deeply-inelastic scattering \cite{MVV}. Details of our calculation are given 
in Ref.~\cite{BKKS}.
\section{Single Scale Feynman Integrals as Recurrent Quantities}
\noindent
For a large variety of massless problems single scale Feynman integrals can be
represented as polynomials in the ring formed of the nested harmonic sums
$S_{a_1, \dots, a_k}(N)$,
\cite{HSUM,SUMMER} and the MZVs $\zeta_{a_1, \dots, a_l}$ over the rational function field ${\bf Q}(N)$.
Here,
\begin{eqnarray}
\label{HS}
S_{b,\vec{a}}(N) = \sum_{k=1}^N \frac{({\rm sign}(b))^k}{k^{|b|}} S_{\vec{a}}(k),\qquad
a_i, b \in {\bf Z} \backslash \{0\},\\
\zeta_{b,\vec{a}} = \sum_{k=1}^\infty\frac{({\rm sign}(b))^k}{k^{|b|}}
S_{\vec{a}}(k),\qquad
a_i, b \in {\bf Z} \backslash \{0\}~.
\nonumber
\end{eqnarray}
If $b=1$, the meaning of $\zeta_{b,\vec{a}}$ is symbolic, since it diverges. The degree
of divergence is a positive power of $S_1(\infty)$.
Rational functions in $N$ and harmonic sums 
obey recurrence relations. Thus, due to closure properties~\cite{GFUN} also any polynomial expression in such terms is a solution of a recurrence.
Consider as an example the recursion
\begin{eqnarray}
F(N+1) - F(N) = \frac{{\rm sign}(a)^{N+1}}{(N+1)^{|a|}}~.
\nonumber
\end{eqnarray}
It is solved by {$S_a(N)$}. Corresponding difference equations hold for
harmonic sums of deeper nestedness.
Feynman integrals can often be decomposed into a combination containing
terms of the  form
\begin{eqnarray}
\int_0^1 dz \frac{z^{N-1}-1}{1-z} H_{\vec{a}}(z)
,~~~~\int_0^1 dz
\frac{(-z)^{N-1}-1}{1+z} H_{\vec{a}}(z)~,
\nonumber
\end{eqnarray}
with $H_{\vec{a}}(z)$ being a harmonic polylogarithm, \cite{VR}.
This structure also leads to recurrences, cf. \cite{STRUCT}. It is very likely
that single scale Feynman diagrams do always obey difference equations.
\section{Establishing and Solving Recurrences}
\noindent
We assume that a sufficiently large set of moments at integer values $N_i$ is given
for a physical quantity, which obeys a recurrence relation. One seeks
\begin{eqnarray}
\sum_{k=0}^l \left[\sum_{i=0}^d c_{i,k} N^i\right] F(N+k) = 0~.
\label{DEQ}
\end{eqnarray}
The method for determining potential recurrences is available in standard
packages \cite{GFUN}.
The corresponding linear system is dense. Rational number arithmetics is not
feasible for the large systems to be solved. Let us consider as an example
the difference equation being associated to the contribution of the color factor
$C_F^3$  for the 3-loop Wilson coefficient  $C_{2,q}^{(3)}$ in unpolarized deeply
inelastic scattering.  {11 Tb} of memory would be required to establish~(\ref{DEQ})
in a naive way.
Therefore refined methods have to be applied. We use arithmetic in {finite fields}
together with Chinese remaindering \cite{CRRR}, which reduces the storage
requirements to a
{few Gb} of memory. The linear system approximately minimizes for {$l \approx d$}.
If one finds more than one recurrence the different recurrences are joined to
reduce {$l$} to a minimal value. It seems to be a general phenomenon that the
recurrence of minimal order is this with the smallest integer coefficients, cf. also
\cite{bostan:08}. For even larger problems than those dealt with in the
present paper, a series of further technical improvements may be carried out,
\cite{TI}.

For the solution of the recurrence {low orders} are clearly preferred.
It is solved in depth-optimal {$\Pi\Sigma$
fields}~\cite{PISIG}; here we apply {advanced symbolic summation} methods as: efficient recurrence solvers and refined telescoping algorithms. They are available in the summation package {\tt Sigma} \cite{SIGMA} implemented in the computer algebra system Mathematica.

The solutions are found as linear combinations of rational terms in {$N$} combined
with functions, which cannot be further reduced in the $\Pi\Sigma$ fields. In the present
application they turn out all to be nested harmonic sums $S_{\vec{b}}(N)$, (\ref{HS}).
Other or higher order applications may lead to sums of {\it different type}
as well, which are uniquely found by the present algorithm.
\section{Determination of the 3-Loop Anomalous Dimensions and Wilson Coefficients}
\noindent
We apply the method to determine the  unpolarized {anomalous dimensions}
and {Wilson coefficients} to {3-loop} order. Here we use the above method to all the
contributions to a single color/$\zeta_i$-factor. These  are 186 terms.
As input to the calculation we use the respective Mellin moments, which were
calculated by a {\tt MAPLE}-code based on the harmonic sum representation \cite{MVV}.
We need very high moments and calculate the input recursively.
As an example, let us illustrate the size of the moments for the
$C_F^3$-contribution to the Wilson coefficient $C_{2,q}(N)$. The highest moment
required is $N = 5114$. It cannot be calculated simply with
{\tt summer} \cite{SUMMER} because of time and storage reasons. The highest
moment is a rational number with a numerator of 13388 and a denominator 13381 digits.
Below we give the moments for $N=3$ and $N=500$.

{\tiny
\begin{verbatim}
N=3:
#11 digits / #10 digits

-98268084191 / 1166400000

N=500:
#1262 digits / #1256 digits

1641840770424196780953020619176376506284303544481262083057197600746507008493793994
4224110323441591630311482222058287688942209570859151121677307585313995100978363179
2518952817622034037186132846974627021672678012913675099511203807811938593043910803
5044345920218696052588332036355325089998361354226882367322149037631053761764348772
5403810874264968729520075619227285471802419403727207822473765999900236383740315299
2050533601633484348249454757555344664210814111140065475391136798689167410065076749
3578709478683390573977410013520894494463909291327425815766566386397276158317387748
5945471392646089700875157445075073192328542890965462004805711998748144414379386093
9937361798029044425789953726133675199790523770427298500510063464061985840066296071
3372543015648919155964069606994597363886301185067827291937065300754786947063672848
9382081926871078600328628131936766057475970450896556667622163365895808773428119721
5352792131089063577045069693962213061198894057033606068695607123271969726981060056
0115846094360239986233917872260722277322690450132376836253549152130116645670565045
9666945920164586023958060271746606798898861360772333088030741775605546518788793327
2264368297071217405654474375844238250889238538974548421298170425909521742559494728
72017877003947396562261659860366839154407853462338171648227013134266795320251847
/
3057444614247225372882570514367358697278130741348282122206492932820352440850471902
7491046962105336645563654873675690796713906565688820365601907263710863954826386081
3227580037879361869941003802807590860358894142891046776447162895908787986423254678
5776778283337231702130612499429819559798501074020676282769289102955679421885795867
1982932998601320344971927374905889934059987271939760212836368619501189238215442366
3805773701929509268157747992859384837403751183019423692868569168206789710047557452
5131217382272060267681480496298975522467614707848639773185909858278799786637303834
1017166676276847525704755493166263297079720470719813623901545811953853986456533543
9994182050551827959988760121168490745476969259468454613431624179198860751513076481
0304734205926703138519418575731315944374873897873646706993620825697218523316375559
4068222004765962715924208526106008109740402380126260947524640509361283802755722132
4856690051525724685919792641506082307567956962328560073471086799287131287564668441
6256698083504233897436484702002471314330803421467773925541151273924985946178771189
2312437162213438137703896064734987157020801413153555435311326719739117599044341913
5922693587373856609594245948237469293148702516714038297077639382332251255360181047
49658623247509112659762997679737527882711111677459300352000000000000000000

N=5114:
#13388 digits / #13381 digits
\end{verbatim}
}

\noindent
The corresponding difference equations (\ref{DEQ}) are determined by a {recurrence
finder}. Furthermore, the order of the difference equation is reduced
to the smallest value possible. The difference equations are then solved order by order
using the summation package {\tt Sigma}.

For the $C_F^3$-term in $C_{2,q}(N)$ the recurrence was established after 20.7 days of
CPU time. Here 4h were required for the modular prediction of the dimension of the
system, 5.8 days were spent on solving modular linear systems, and 11 days for the
modular operator GCDs. The Chinese remainder method and rational reconstruction
took 3.8 days. 140 word size primes were needed. As output one obtains a
recurrence of 31 Mb, which is of
order 35 and degree 938, with a largest integer of 1227 digits.
The recurrence was solved by {\tt Sigma} after 5.9 days. We reached a
compactification
from 289 harmonic sums needed in \cite{MVV} to 58 harmonic sums, where the
representation in \cite{MVV} (see a corresponding attachment) has already been
compactified following an idea of one of the present authors. The determination of the
3-loop anomalous dimensions is a  much smaller problem. Here the computation
takes about 18~h only for the complete result.

For the three most complicated cases,
establishing and solving of the difference equations took $3+1$ weeks each, requiring
$\leq 10$Gb on a 2~GHz processor. This led to an overall computation time of
about sixteen weeks, with the possibility to parallelize four times. Here we did not yet
consider parallelization w.r.t. the 140 primes chosen, which would significantly reduce the
computational time of the $C_F^3$ term discussed above and for other comparably large
contributions.

In the final representation, we account for algebraic reduction \cite{ALGEBRA}; for this task we used the package {\tt HarmonicSums}~\cite{Ablinger:09} which complements the functionalities of~{\tt Sigma}. One observes that
different color
factor contributions lead to the same, or nearly the same, amount of sums at a
given quantity. This points to the fact that the amount of sums contributing,
after the algebraic reduction has been carried out, is governed
by topology rather than the field- and color structures being involved. The linear harmonic sum
representations used in \cite{MVV} require many more sums than in the
representation reached by the present analysis. A further reduction can be obtained using
the \emph{structural
relations}, which leads to maximally {35 different sums} up to the level the
3-loop Wilson coefficients \cite{STRUCT}. It is not unlikely that the present 
method can be applied to single scale problems in even higher order. As has 
been found
before in \cite{STRUCT,TWL,HEAV,EPEM} representing a large number
of 2- and 3-loop processes in terms of harmonic sums, the {basis elements} emerging
are always the same. This applies to the anomalous dimensions and Wilson coefficients of
the space- and time-like  polarized and unpolarized case, the polarized and
unpolarized Drell-Yan process and hadronic Higgs-Boson production in the heavy mass limit,
deep-inelastic heavy flavor production in the limit $Q^2 \gg m_Q^2$,
higher order QED  corrections in $e^+e^-$ annihilation, as well as to soft and virtual
corrections to Bhabha scattering.

In practice no method does yet exist to calculate such a high number of moments
ab initio as required for the determination of the all $N$ formulae in the 3--loop case.
On the other hand,
a proof of existence has been delivered of a quite general and powerful automatic
difference-equation solver, standing rather demanding tests, which is ready to help
in the solution of present day problems in higher order Quantum Field Theory. It opens
up good prospects for the development of even more powerful methods.
\section{Conclusions}

\noindent
We established a general algorithm to calculate the exact expression
for single scale quantities from a finite (suitably large) number of moments,
which are zero scale quantities. The latter ones are much more easily calculable
than single scale quantities.
We applied the method to the anomalous dimensions and Wilson coefficients up to
3-loop order. To solve 3-loop problems this way is not possible at present,
since the number of required moments is too large for the methods available.
We established and solved the recurrences for all color resp. $\zeta$-projections
at once, which forms a rather voluminous problem. Yet we showed that giant difference
equations [order 35;
degree $\sim$ 1000] can be reliably and fast established and solved unconditionally
for the most advanced problems in Quantum Field Theory.

\vspace{3mm}\noindent
{\bf Acknowledgment.}~ We would like to thank J. Vermaseren for discussions.


\begin{thebibliography}{99}
\bibitem{LIS}
  J.~M.~Borwein, D.~M.~Bradley, D.~J.~Broadhurst and P.~Lisonek,
  Trans.\ Am.\ Math.\ Soc.\  {\bf 353} (2001) 907
  [arXiv:math/9910045].
\bibitem{BBV}
J. Bl\"umlein, D. Broadhurst, J. Vermaseren, \emph{The multiple zeta value data
mine}, DESY 09-003.
\bibitem{ANDRE}
  D.~J.~Broadhurst,
  Eur.\ Phys.\ J.\  C {\bf 8} (1999) 311
  [arXiv:hep-th/9803091];\\
F. Brown, arxiv:0804.1660 [math.AG];\\
Y. Andre, Proceedings of the International Conference ``Motives, Quantum Field
Theory, an Pseudo Differential Operators'', Clay Mathematical Institute,
Boston,
June, 2008.
\bibitem{PSLQ}
H.R.P. Ferguson and D.H. Bailey, D. H. \emph{A Polynomial Time, Numerically Stable
Integer Relation Algorithm}, RNR Techn. Rept. RNR-91-032, Jul. 14, 199.
\bibitem{BROAD96}
  D.~J.~Broadhurst,
  {arXiv:hep-th/9604128}.
\bibitem{BKZ}
  D.~J.~Broadhurst and D.~Kreimer,
  Phys.\ Lett.\  B {\bf 393} (1997) 403,
  {[arXiv:hep-th/9609128]};\\
D. Zagier,
in~: First European Congress
of Mathematics, Vol. II, (Paris, 1992), Progr. Math., {\bf 120}, (Birkh\"auser, Basel--Boston, 1994),
pp.~497.
\bibitem{TERAS}
A.B. Goncharov,
{arxiv:math.AG/0103059};\\
T. Terasoma,
Invent. Math. {\bf 149} (2) (2002) 339, {arxiv:math.AG/010423};\\
P. Deligne and A.B. Goncharov,
Ann. Sci. Ecole Norm. Sup., S\'erie IV {\bf 38} (1) (2005) 1.
\bibitem{MVV}
  S.~Moch, J.~A.~M.~Vermaseren and A.~Vogt,
  Nucl.\ Phys.\  B {\bf 688} (2004) 101
  [arXiv:hep-ph/0403192];
  Nucl.\ Phys.\  B {\bf 691} (2004) 129
  [arXiv:hep-ph/0404111];
  Nucl.\ Phys.\  B {\bf 724} (2005) 3
  [arXiv:hep-ph/0504242].
\bibitem{BKKS}
J. Bl\"umlein, M. Kauers, S. Klein, and C. Schneider, {arxiv: 0902.4091 [hep-ph]}.
%
\bibitem{HSUM}
  J.~Bl\"umlein and S.~Kurth,
  Phys.\ Rev.\  D {\bf 60} (1999) 014018
  [arXiv:hep-ph/9810241];\\
\bibitem{SUMMER}
  J.~A.~M.~Vermaseren,
  Int.\ J.\ Mod.\ Phys.\  A {\bf 14} (1999) 2037
  [arXiv:hep-ph/9806280].
\bibitem{VR}
  E.~Remiddi and J.~A.~M.~Vermaseren,
  Int.\ J.\ Mod.\ Phys.\  A {\bf 15} (2000) 725--754,
  {[arXiv:hep-ph/9905237].}
\bibitem{STRUCT}
  J.~Bl\"umlein,
  Nucl.\ Phys.\ Proc.\ Suppl.\  {\bf 183} (2008) 232
  [arXiv:0807.0700 [math-ph]];
{arxiv:0901.0837}, {arxiv:0901.3106}.
\bibitem{GFUN}
B. Salvy and P. Zimmermann,
ACM Transactions on Mathematical Software, {\bf 20} (2) (1994) 163;\\
C. Mallinger,
Master Thesis, J. Kepler University, Linz, (1996).
\bibitem{CRRR}
K.O. Geddes, S. R. Czapor and G. Labahn,
\emph{Algorithms for Computer Algebra}, (Kluwer, Dordrecht, 1992);\\
J. von zur Gathen and J. Gerhard,
\emph{Modern Computer Algebra}, (Cambridge University Press, Cambridge,
1999);\\
M. Kauers,
Nucl. Phys. {\bf B} (Proc. Suppl.) {\bf 183} (2008) 245;\\
%
\bibitem{bostan:08}
A. Bostan and M. Kauers,
\emph{The full counting function for {G}essel walks is algebraic},
INRIA-Rocquencourt report, 2009,
in preparation.
%
\bibitem{TI}
B. Beckermann and G. Labahn,
Numerical Algorithms, {\bf 3} (1992) 45;
SIAM Journal of Matrix Analysis and Applications,
{\bf 22} (1) (2000) 114.
\bibitem{PISIG}
M.~Karr,
{J.~ACM}, {\bf 28} (1981) 305;
{J.~Symbolic Comput.}, {\bf 1} (1985) 303;\\
C.~Schneider,
PhD thesis, RISC-Linz, J.~Kepler University, Linz, May 2001;
{J. Differ. Equations Appl.}, {\bf 11}(9) (2005) 799;
{J. Algebra Appl.}, {\bf 6} (3) (2007) 415;
\emph{Symbolic summation finds optimal nested sum representations}
SFB-Report 2007-26, SFB F013, J. Kepler University Linz, 2007;
{\it Parameterized telescoping proves algebraic independence of sums}
{Ann. Comb.}, to appear, 2008;
{J. Symbolic Comput.}, {\bf 43} (9) (2008) 611.
\bibitem{SIGMA}
C.~Schneider,
{S\'em.~Lothar. Combin.}, {\bf 56} (2007) 1, Article B56b.
\bibitem{ALGEBRA}
  J.~Bl\"umlein,
  Comput.\ Phys.\ Commun.\  {\bf 159} (2004) 19
  [arXiv:hep-ph/0311046].
\bibitem{Ablinger:09}
J.~Ablinger.
Diploma thesis, J. Kepler University Linz, 2009.
\bibitem{TWL}
  J.~Bl\"umlein and V.~Ravindran,
  Nucl.\ Phys.\ B {\bf 716} (2005) 128
  [arXiv:hep-ph/0501178];
  Nucl.\ Phys.\ B {\bf 749} (2006) 1
  [arXiv:hep-ph/0604019];
\\
J. Bl\"umlein and S. Moch, in preparation;\\
  J.~Bl\"umlein and S.~Klein,
  arXiv:0706.2426 [hep-ph].
\bibitem{HEAV}
  J.~Bl\"umlein, A.~De Freitas, W.~L.~van Neerven and S.~Klein,
  Nucl.\ Phys.\ B {\bf 755} (2006) 272
  [arXiv:hep-ph/0608024];\\
  M.~Buza, Y.~Matiounine, J.~Smith, R.~Migneron and W.~L.~van Neerven,
  Nucl.\ Phys.\ B {\bf 472} (1996) 611
  [arXiv:hep-ph/9601302];\\
  I.~Bierenbaum, J.~Bl\"umlein, S.~Klein and C.~Schneider,
  Nucl.\ Phys.\  B {\bf 803} (2008) 1
  [arXiv:0803.0273 [hep-ph]];\\
  I.~Bierenbaum, J.~Bl\"umlein and S.~Klein,
  Nucl.\ Phys.\  B {\bf 780} (2007) 40
  [arXiv:hep-ph/0703285];
{arxiv: 0901.0669}, Phys. Lett. {\bf B} (2009) in print.
\bibitem{EPEM}
  F.~A.~Berends, W.~L.~van Neerven and G.~J.~H.~Burgers,
  Nucl.\ Phys.\  B {\bf 297} (1988) 429
  [Erratum-ibid.\  B {\bf 304} (1988) 921];\\
  J.~Bl\"umlein, A.~De Freitas, W.~van Neerven,
  PoS \emph{RADCOR2007} (2007) 005
  [arXiv:0812.1588 [hep-ph]].
\end{thebibliography}
\end{document}